\documentclass[11pt]{article}

\usepackage[letterpaper, left=1in, top=1in, right=1in, bottom=1in,nohead,includefoot]{geometry}
\usepackage{color,charter,graphicx,natbib,here,setspace,multirow,amsmath,amssymb,amsfonts,enumitem}
\usepackage[dvipsnames]{xcolor}
\usepackage{graphicx}
\usepackage{rotating}
\usepackage{multirow}
\usepackage{booktabs}
\usepackage{color}
\usepackage{setspace}
\usepackage{algorithm2e}
\usepackage{enumitem}
\usepackage[bottom]{footmisc}
\usepackage[colorlinks=true,urlcolor=blue,linkcolor=blue,citecolor=blue]{hyperref}
\usepackage{bigstrut}
\usepackage{threeparttable}
\usepackage{dcolumn}
\usepackage{amsmath}
\usepackage{longtable}
\usepackage{amssymb}
\usepackage{bm}
\pdfoutput=1
\usepackage{eqparbox}

\usepackage[title,titletoc]{appendix}
\makeatletter

\renewenvironment{appendices}{%
    \begin{oldappendices}%
    \renewcommand{\thefigure}{\ifnum \c@section>\z@ \thesection.\fi\@arabic\c@figure}%
    \@addtoreset{figure}{section}%
    \renewcommand{\thetable}{\ifnum \c@section>\z@ \thesection.\fi\@arabic\c@table}%
    \@addtoreset{table}{section}%
}{%
    \end{oldappendices}%
}\makeatother

\usepackage{natbib}
\let\natbibcitet\citet
\renewcommand\citet{\bibpunct{(}{)}{,}{a}{,}{,}\natbibcitet}
\let\natbibcitep\citep
\renewcommand\citep{\bibpunct{(}{)}{;}{a}{,}{;}\natbibcitep}
\newcommand{\bi}{\begin{itemize}}
\newcommand{\ei}{\end{itemize}}
\newcommand{\be}{\begin{equation}}
\newcommand{\ee}{\end{equation}}
\defcitealias{ieo14}{IEO, 2014}

\long\def\symbolfootnote[#1]#2{\begingroup%
\def\thefootnote{\fnsymbol{footnote}}\footnote[#1]{#2}\endgroup}

\usepackage{sectsty}
\allsectionsfont{\normalsize}

\makeatletter

\@addtoreset{equation}{section}
\makeatother

\usepackage{caption}
\usepackage[labelformat=simple]{subcaption}
\makeatletter\let\p@subfigure\thefigure\makeatother

\usepackage{cleveref}
\crefname{chapter}{Chapter}{Chapters}
\crefname{section}{Section}{Sections}
\crefname{subsection}{Section}{Sections}
\crefname{subsubsection}{Section}{Sections}
\crefname{figure}{Figure}{Figures}
\crefname{table}{Table}{Tables}
\crefname{equation}{Equation}{Equations}
\crefname{appendix}{Appendix}{Appendices}

\newcolumntype{d}[1]{D{.}{.}{#1}}

\usepackage{authblk}

  \title{Introducing shrinkage in heavy-tailed  state space models to\\[-.5em] predict equity excess returns}
  \author[1]{Florian Huber}
  \author[2]{Gregor Kastner\thanks{Corresponding author. Address: Institute for Statistics and Mathematics, Department of Finance, Accounting and Statistics, WU Vienna University of Economics and Business. Welthandelsplatz 1 / D4 / level 4, 1020 Vienna, Austria. Phone: +43 1 31336-5593. E-mail: \href{mailto:gregor.kastner@wu.ac.at}{gregor.kastner@wu.ac.at}. The authors acknowledge funding from the Austrian Science Fund (FWF) for the project ``High-dimensional statistical learning: New methods to advance economic and sustainability policies'' (ZK 35), jointly carried out by WU Vienna University of Economics and Business, Paris Lodron University Salzburg, TU Wien, and the Austrian Institute of Economic Research (WIFO).}}
  \author[1]{Michael Pfarrhofer}
  \affil[1]{Salzburg Centre of European Union Studies, University of Salzburg}
  \affil[2]{Department of Finance, Accounting and Statistics, WU Vienna}
  

\def\equationautorefname~#1\null{%
  Eq.~(#1)\null
}
\def\equationautorefname~#1\null{
Eq.~(#1)\null
}

\begin{document}
\doublespacing
\graphicspath{{Figs/}}
 \maketitle

 \begin{abstract}\noindent
We forecast S\&P 500 excess returns using a flexible Bayesian econometric state space model with non-Gaussian features at several levels. More precisely, we control for overparameterization via novel global-local shrinkage priors on the state innovation variances as well as the time-invariant part of the state space model. The shrinkage priors are complemented by heavy tailed state innovations that cater for potential large breaks in the latent states. Moreover, we allow for leptokurtic stochastic volatility in the observation equation. The empirical findings indicate that several variants of the proposed approach outperform typical competitors frequently used in the literature, both in terms of point and density forecasts.
 \end{abstract}


\section{Introduction}
\label{intro}
Predicting equity prices has been one of the main challenges for financial economists during the last decades. A plethora of studies emerged that draw a relationship between different macroeconomic and financial fundamentals and the predictability of excess returns \citep{lettau2001consumption, ang2007stock, welch2008comprehensive, dangl2012predictive}. While some authors find evidence of predictability, simple naive benchmarks still prove to be extremely difficult to beat by more sophisticated models.

In this paper, we aim to predict S\&P 500 excess returns by proposing a flexible dynamic regression model. \cite{dangl2012predictive} postulate a time-varying relationship between excess returns $y_t$ and a set of $K$ fundamental predictors in $\bm{X}_t$, given by the dynamic regression model
\begin{align}
y_t &= \bm{\beta}'_t \bm{X}_t + \varepsilon_t,\label{eq: obs} \\ 
\bm{\beta}_t&=\bm{\beta}_{t-1}+\bm{w}_t, \label{eq: states}
\end{align}
for $t = 1,\dots, T$ \citep{west2006bayesian}. Here, it is assumed that the regressors are related to $y_t$ through a set of $K$ dynamic (time-varying) regression coefficients $\bm{\beta}_t$ that follow a random walk process with $\bm{w}_t \sim \mathcal{N}(\bm{0}_K, \bm{V})$, where $\bm{V}=\text{diag}(v_1, \dots, v_K)$ is a diagonal variance-covariance matrix of dimension $K\times K$. To simplify computation,  the measurement errors captured through $\varepsilon_t$ are often assumed to follow a zero mean Gaussian distribution with variance $\sigma^2_\varepsilon$.

Model specification within this econometric framework received considerable attention recently \citep[see, among many others,][]{fruhwirth2010stochastic, eisenstat2016stochastic, bitto2015achieving}. One prevalent issue is that, if left unrestricted, \autoref{eq: obs} has a strong tendency to overfit the data, leading to imprecise out-of-sample forecasts. This calls for some form of regularization. \cite{fruhwirth2010stochastic} show how a non-centered parameterization of the state space model can be used to apply a standard Bayesian shrinkage prior on the process variances in $\bm{V}$. This approach allows for capturing model uncertainty along two dimensions: The first dimension is whether a given element in $\bm{X}_t$, $X_{jt}$, should be included or excluded. The second dimension addresses the question whether the associated element in $\bm{\beta}_t$, $\beta_{jt}$, should be constant or time-varying. Note that the latter is equivalent to setting $v_j=0$, which yields $\beta_{jt}=\beta_{jt-1}$ for all $t$. 

In the present contribution we combine the literature on shrinkage and variable selection within the general class of state space models \citep{fruhwirth2010stochastic, eisenstat2016stochastic, bitto2015achieving} with the literature on non-Gaussian state space models \citep{carlin1992monte, kitagawa1996monte}. The model we propose features t-distributed shocks to both the observation and the state equation. This choice provides enough flexibility to capture large outliers commonly observed in stock markets. To cope with model and specification uncertainty, we adopt the Dirichlet-Laplace \citep[DL,][]{bhattacharya2015dirichlet} shrinkage prior that allows for flexible shrinkage towards simpler nested model specifications. One empirical key observation from the macroeconomic literature \citep{sims2006were, koop2009evolution} is that parameters tend to change abruptly, as opposed to smoothly. We capture this stylized fact by assuming that the shocks to the states follow a (potentially) heavy tailed t-distribution that allows for large jumps in the regression coefficients, even in the presence of strong shrinkage towards constancy.

To investigate whether these extensions translate to predictive gains, we apply our proposed model framework to the well-known dataset compiled in \cite{welch2008comprehensive}.  More specifically, we forecast monthly S\&P~500 excess returns over a period of 55 years and compute one-step-ahead predictive densities.\footnote{S\&P~500 data is available at a number of different frequencies. For the purpose of this paper and limited by the availability of higher-frequency covariates, we opt for monthly observations.} We then assess to what extent the proposed methods outperform simpler nested alternatives and other competing approaches both in terms of root mean square errors (RMSEs) as well as log predictive scores (LPS).

Our results indicate that a time-varying parameter model with shrinkage and heavy tailed measurement errors displays the best predictive performance over the full hold-out period. Considering the results within expansions and recessions highlights that allowing for heavy tailed state innovations pays off in economic downturns, while it is outperformed by a specification with heavy tailed measurement errors in expansions. A dynamic model selection exercise shows that forecasting performance may be further improved by computing model weights based on previous predictive likelihoods. Strong overall forecasts generally translate into a favorable performance in terms of Sharpe ratios. Using this economic evaluation criterion suggests that models that work well in forecasting also work well when used to generate trading signals.

The remainder of the paper is structured as follows. Section \ref{sec: ectrcs} introduces the necessary modifications to the econometric model postulated in Eqs.\ (\ref{eq: obs}) and (\ref{eq: states}) to allow for heavy-tailed measurement and state innovations. In addition, the section provides an overview on the Bayesian prior setup and a brief sketch of the Markov chain Monte Carlo (MCMC) algorithm. Section \ref{sec:empirics} presents the empirical results, focusing first on selected in-sample features of the model before discussing the results of our forecasting exercise. Finally, the last section summarizes and concludes the paper.

\section{Econometric framework}\label{sec: ectrcs}
\subsection{A non-Gaussian state space model}
In Section~\ref{intro}, the shocks to both the measurements as well as the states are assumed to follow Gaussian distributions with constant variances. For financial data, however, this could be overly restrictive and especially the assumption of homoscedasticity is likely to translate into weak density forecasts.

As a remedy, we propose the measurement errors to follow a t-distribution with $\nu$ degrees of freedom and time-varying variance,
\begin{align}
\varepsilon_t|h_t, \nu &\sim {t}_\nu(0, e^{h_t}),\label{eq: tinnovmeas}\\
h_t|h_{t-1} &\sim \mathcal{N}(\mu + \rho (h_{t-1}-\mu), \sigma^2_h),\\
h_0 &\sim \mathcal{N}\left(\mu, \sigma_h^2/(1-\rho^2)\right),
\end{align}
where $\mu$ denotes the unconditional mean of the log-volatility process $h_t$, $\rho$ its autoregressive parameter and $\sigma^2_h$ its innovation variance.  Introducing auxiliary variables $\bm{\tau} = (\tau_{1}, \dots, \tau_T)'$ permits stating \autoref{eq: tinnovmeas} as a conditional Gaussian distribution,
\begin{align}
\varepsilon_t|h_t,\tau_{t} &\sim \mathcal{N}(0, \tau_{t} e^{h_t}), \label{eq: tinnovmeas_GAUSS}\\
\tau_t &\sim \mathcal{G}^{-1}(\nu/2, \nu/2).
\end{align}
This specification of the measurement errors allows to capture large shocks as well as time-variation in the underlying error variances. Especially for financial data that are characterized by heavy tailed shock distributions as well as heteroscedasticity, this proves to be a key feature to produce precise predictive densities.

Furthermore, we assume that the shocks to the latent states follow a heavy tailed error distribution. Similarly to \autoref{eq: tinnovmeas} and \autoref{eq: tinnovmeas_GAUSS}, the state innovations follow a t-distribution with $\kappa_j$ degrees of freedom,
\begin{align}
w_{jt}|\kappa_j \sim t_{\kappa_j}(0, v_j)\quad \Leftrightarrow\quad w_{jt}|\xi_{jt} \sim \mathcal{N}(0, \xi_{jt} v_j), \label{eq: shocks_states}
\end{align}
where the elements of $\bm{\xi}_j = (\xi_{j1},\dots,\xi_{jT})$ follow independent $\mathcal{G}^{-1}(\kappa_j/2, \kappa_j/2)$ distributions. In contrast to \autoref{eq: tinnovmeas}, we assume that the shocks to the states are homoscedastic. Notice that \autoref{eq: shocks_states} effectively implies that we occasionally expect larger breaks in the underlying regression coefficients, even if $v_j$ is close to zero. This appears to be of particular importance when shrinkage priors are placed on $v_j$.

\subsection{A Dirichlet-Laplace shrinkage prior}
The model described in the previous sections is heavily parameterized and calls for some sort of regularization in order to provide robust and accurate forecasts. To this end, we follow \cite{fruhwirth2010stochastic} and exploit the non-centered parameterization of the model,
\begin{align}
y_t =& \bm \beta_0' \bm X_t + \tilde{\bm \beta}'_t \sqrt{\bm V} \bm Z_t + \varepsilon_t, \label{eq: NCP_OBS}\\
\tilde{\bm \beta}_t =& \tilde{\bm \beta}_{t-1} + \bm \eta_t,\quad \bm \eta_t \sim \mathcal{N}(\bm 0_K, \bm I_K).
\end{align}
The $j$th element of $\tilde{\bm \beta}_t$ is given by $\tilde{\beta}_{jt}=\frac{\beta_{jt}-\beta_{j0}}{\xi_{jt}\sqrt{v_{j}}}$, $\bm V = \sqrt{\bm V} \sqrt{\bm V}$, and $\bm Z_t$ is a $K$-dimensional vector with $j$th element given by $Z_{jt} = \sqrt{\xi_{jt}} X_{jt}$. For identification, we set $\tilde{\bm{\beta}}_0 = \bm 0$. Notice that \autoref{eq: NCP_OBS} implies that the process innovation variances as well as the auxiliary variables are transformed from the state to the observation equation. We exploit this by estimating the elements of $\bm \beta_0$ and $\sqrt{\bm V}$ through a standard Bayesian regression model.

We use a Dirichlet-Laplace shrinkage prior \citep{bhattacharya2015dirichlet} on  $\bm \alpha = (\bm \beta_0', \sqrt{v_1}, \dots, \sqrt{v_K})'$. More specifically, for each of the $2K$ elements of $\bm \alpha$, denoted by $\alpha_j$, we impose a hierarchical Gaussian prior given by
\begin{equation}
\alpha_j \sim \mathcal{N}(0, \psi_j \phi^2_j  \lambda^2), \quad \psi_j \sim Exp(1/2), \quad \phi_j \sim Dir(a, \dots, a), \quad \lambda \sim \mathcal{G}(2 K a, 1/2).
\end{equation}
Here, $\psi_j$ denotes a local scaling parameter that is equipped with an exponentially distributed prior and $\bm \phi = (\phi_1, \dots, \phi_{2K})$ is a vector of additional scaling parameters that are restricted to the $(2K-1)$-dimensional simplex, i.e. $\phi_j>0$ for all $j$ and $\sum_{j=1}^{2K} \phi_j = 1$. For each $\phi_j$, we assume a symmetric Dirichlet distribution with intensity parameter $a$ which we set to $a=1/(2K)$ in the empirical application.\footnote{For a theoretical discussion of this choice, see \cite{bhattacharya2015dirichlet}.}  Finally, we let $\lambda$ denote a global shrinkage parameter that pulls all elements in $\bm \alpha$ to zero. Due to the importance of this scaling parameter, we do not fix it a priori but impose a Gamma hyperprior and subsequently infer it from the data. 

\begin{figure}[t]
\includegraphics[width = .49\textwidth]{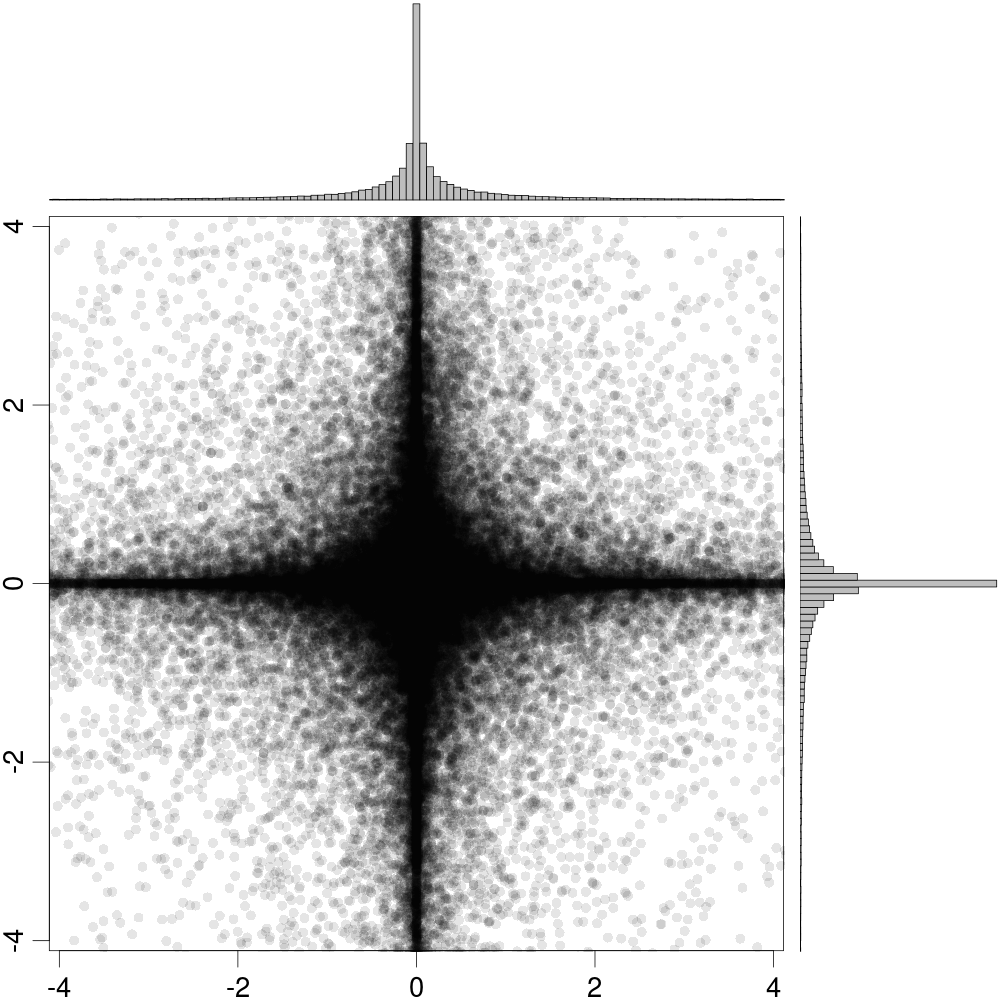}\hfill
\includegraphics[width = .49\textwidth]{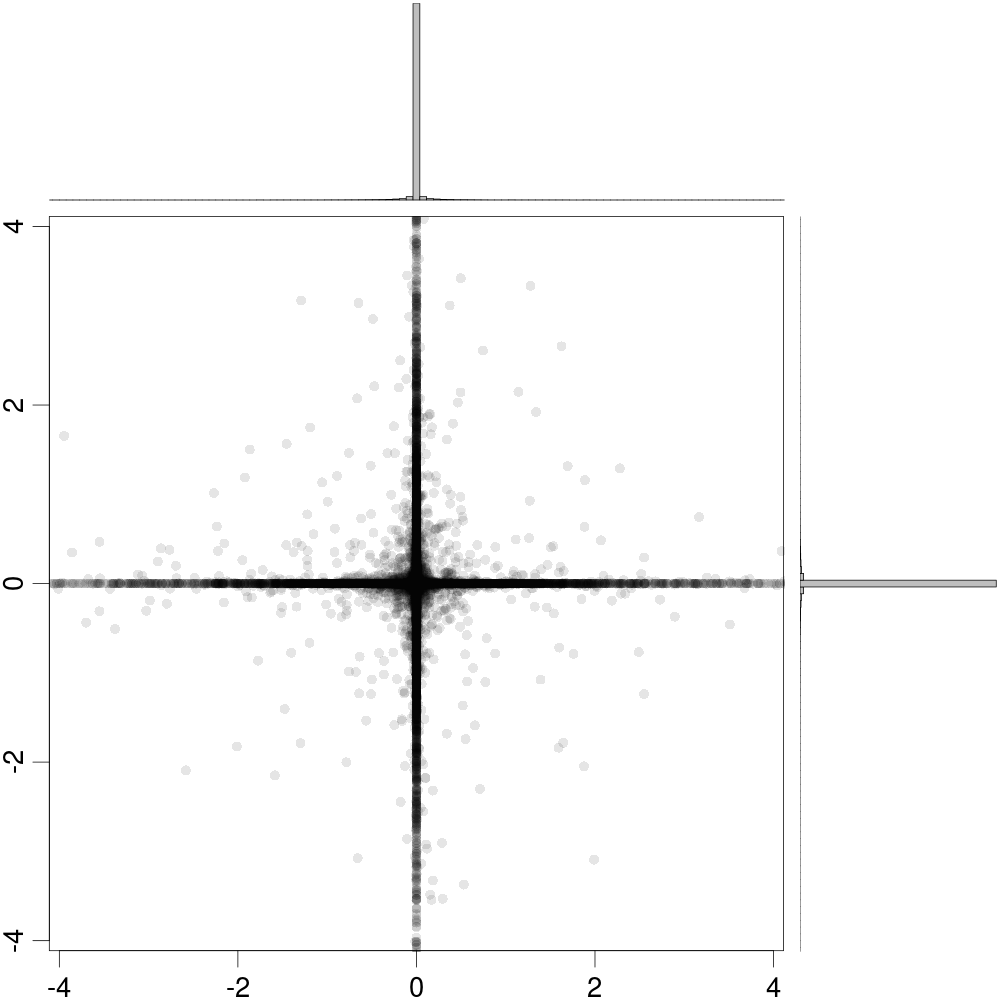}
\caption{Scatterplots and histograms of $100\,000$ draws from the first two components of a DL($1/(2K)$) prior for $K = 1$ (left) and $K = 13$ (right).}
\label{fig:priorsim}
\end{figure}

This prior setup has been shown to perform well for different models and applications \citep[e.g.][]{li2017variable, feldkircher2017sophisticated, kas-hub:spa}. Intuitively, it mimics the behavior of a point mass mixture prior but with the main advantage of computational tractability in high dimensions. The underlying marginal priors on $\alpha_j$ are all heavy tailed, implying that even in the presence of a small global shrinkage parameter $\lambda$, we still allow for non-zero elements in $\bm \alpha$. This feature has been identified to be crucial for good forecasting performance and, in addition, does well in discriminating signals from noise. In \autoref{fig:priorsim}, the first two components of this prior are visualized for a univariate ($K = 1$) and a multivariate dynamic regression setting with $K = 13$ as in the empirical study in Section~\ref{sec:empirics}. Note that both the dependence structure as well as the marginal shrinkage effect becomes stronger with increasing $K$, while the kurtosis remains relatively stable.

This prior introduces shrinkage on the square root of the process innovation variances. Thus, we effectively assess whether coefficients are constant or time-varying within a unified modeling framework.\footnote{For a recent application of shrinkage priors to state space models, see \cite{bitto2015achieving}.} One key advantage of our model, however, is that the heavy tailed innovations allow for breaks in the parameters even if the corresponding process innovation variances are close to zero. Thus, our framework is able to mimic models that only assume a small number of breaks in the regression coefficients, if necessary. 

For the remaining coefficients, we follow \cite{kim-etal:sto} and \cite{kastner2014ancillarity} and use a mildly informative Gaussian prior on the level of log variance, $\mu \sim \mathcal{N}(0, 10^2)$. On the (transformed) persistence parameter we use a Beta prior $\frac{\rho +1}{2} \sim \mathcal{B}(25, 5)$ and on $\sigma^2_h$ we use a Gamma prior, $\sigma^2_h \sim \mathcal{G}(1/2, 1/2)$. Finally, on the degrees of freedom $\nu$ and $\kappa_j$ we impose independent $\mathcal{G}(1, 1/10)$ priors implying that both the prior means as well as the prior standard deviations are equal to $10$.\footnote{To avoid draws that imply infinite conditional variance or ``almost-Gaussianity'' we furthermore restrict the degrees of freedom to the interval $[2,50]$. This particular choice, however, shows almost no influence on the results reported in Section~\ref{sec:empirics}.}

\subsection{Full conditional posterior simulation}
We carry out posterior inference using an MCMC algorithm that is repeated $30\,000$ times with the first $15\,000$ draws discarded as burn-in. The full conditional posterior distributions all have well-known forms, and we can thus set up a Gibbs sampling algorithm that iteratively draws from all relevant distributions. Considered individually, each step has been discussed in previous papers. We provide a brief summary below.
\begin{itemize}
\item Conditional on the remaining parameters and states, we simulate the full history of $\tilde{\bm \beta_t}$ for $t = 1,\dots,T$ using a standard forward filtering backward sampling (FFBS) algorithm \citep{carter1994gibbs, fruhwirth1994data}.
\item  $\bm \beta_0$ as well as the diagonal elements of $\sqrt{\bm V}$ are simulated from a Gaussian conditional posterior distribution by noting that \autoref{eq: NCP_OBS} resembles a standard regression model with heteroscedastic shocks.
\item The full conditional distribution of the local shrinkage parameters is inverse Gaussian, i.e.\ 
$
   \psi_j|\bullet \sim {iG}(\phi_j \lambda/ |\alpha_j|, 1), \ j=1,\dots,2K
   $. To draw from this distribution, we use the rejection sampler of \citet{hoermann2013generating} via the R package \texttt{GIGrvg} \citep{r:gig}.
\item The global shrinkage parameter conditionally follows a generalized inverse Gaussian distribution, i.e.\
 $
 \lambda|\bullet \sim \mathcal{GIG}\left(2K(a - 1),1, 2\sum_{j=1}^{2K}  |\alpha_j|/\phi_j\right)
$, which is again easily accessible through \texttt{GIGrvg}.

\item The scaling parameters $\phi_j$ are drawn by  first sampling auxiliary quantities $T_j$ from
$
\mathcal{GIG}(a-1, 1, 2|\alpha_j|),
$
and then setting 
$
\phi_j = T_j/\sum_{i=1}^{2K} T_i 
$ which yields a draw from $\bm{\phi}|\bm{\alpha}$ \citep{bhattacharya2015dirichlet}.

\item Each element of the auxiliary vector $\bm{\tau}$ is conditionally inverse Gamma distributed, i.e.\ 
$
\tau_t|\bullet \sim \mathcal{G}^{-1}\{(\nu + 1)/2,(\nu + \epsilon_t^2\exp(-h_t))/2\},
$
independently for $t \in \{1, \dots, T\}$, which makes sampling from this distribution straightforward. Draws from $\bm{\xi}_j|\bullet$ for all $j$ are obtained analogously.

\item The conditional likelihood for the degrees of freedom parameter $\nu$ reads
\begin{eqnarray}
 \label{dfdens}
p(\bm{\tau}|\nu) \propto
\left(\frac{\nu}{2}\right)^{n\nu/2}
\Gamma\!\left(\frac{\nu}{2}\right)^{-n}
\left(\prod_{t=1}^n \tau_t\right)^{-\nu/2}
\exp\left\{-\frac{\nu}{2}\sum_{t=1}^n\frac{1}{\tau_t}\right\}.
\end{eqnarray}
To obtain draws from the full conditional distribution, $\nu|\bullet = \nu|\bm{\tau}$, we use an independence Metropolis-Hastings update in the spirit of \cite{chi-gre:bay}. We find the maximizer of \autoref{dfdens} and the corresponding Fisher information which we, in turn, use to construct a Gaussian proposal distribution. For details, see \citet{hos-kas:mod} and \citet{kas:hea}. Draws from $\kappa_j|\bullet$ for all $j$ are obtained analogously.

\item Conditional on all other parameters, updating the latent log variances $\bm{h} = (h_0, h_1, \dots, h_T)$ and the stochastic volatility parameters $\mu$, $\rho$, and $\sigma_h^2$ is done exactly as in \cite{kastner2014ancillarity}, who utilize an efficient auxiliary mixture sampler \citep{omo-etal:sto} with ancillarity-sufficiency interweaving \citep[ASIS,][]{yu-men:cen}. We access this sampler through the implementation in the R package \texttt{stochvol} \citep{Kastner2016dealing}.
\end{itemize}

\section{Empirical application}
\label{sec:empirics}
In this section we start by providing information on the data and model specification in Section~\ref{sec:data}, followed by key empirical findings of our model for the full sample period in Section~\ref{sec:insample}. We proceed by describing the forecasting design and the set of competing models in Section~\ref{sec:design}. The main forecasting results are presented in Section~\ref{sec:forecast}.

\subsection{Data overview and model specification}
\label{sec:data}
We adopt the dataset utilized in \cite{welch2008comprehensive} and establish a relationship between S\&P~500 excess returns and a set of fundamental factors that are commonly used in the literature. Our dataset is monthly and spans the period from 1927:01 to 2010:12. The response variable is the S\&P~500 index return minus the risk free rate. 

The following lagged explanatory variables are included in our models: The dividend price ratio (DP), the dividend yield (DY), the earnings price ratio (EP), the stock variance (SVAR, defined as the sum of squared S\&P~500 daily returns), and the book-to-market ratio (BM). Furthermore, we include the ratio of 12-month moving sums of net issues by stocks listed at the New York Stock Exchange (NYSE) divided by the total end-of-year market capitalization of NYSE stocks (NTIS). Moreover, the models feature yields on short- and long-term government debt and information on term spreads (TBL, LTY and LTR). To capture corporate bond market dynamics, we rely on the spread differences between BAA and AAA rated corporate bond yields and the differences of corporate and treasury bond returns at the long end of the yield curve (DFY and DFR). Finally, the set of covariates is completed by consumer price inflation (INFL) and an intercept term (cons). For more information on the construction of the exogenous variables that mainly capture stock characteristics, see \cite{welch2008comprehensive}.

\subsection{Empirical results for the full sample period}\label{sec:insample}
In this section, we use our proposed non-Gaussian state space model to provide some evidence for time variation in the coefficients of the model. We first focus on the measurement errors, and subsequently extend the discussion to the time varying regression coeffients.

\begin{figure}[t]
\includegraphics[width=\textwidth]{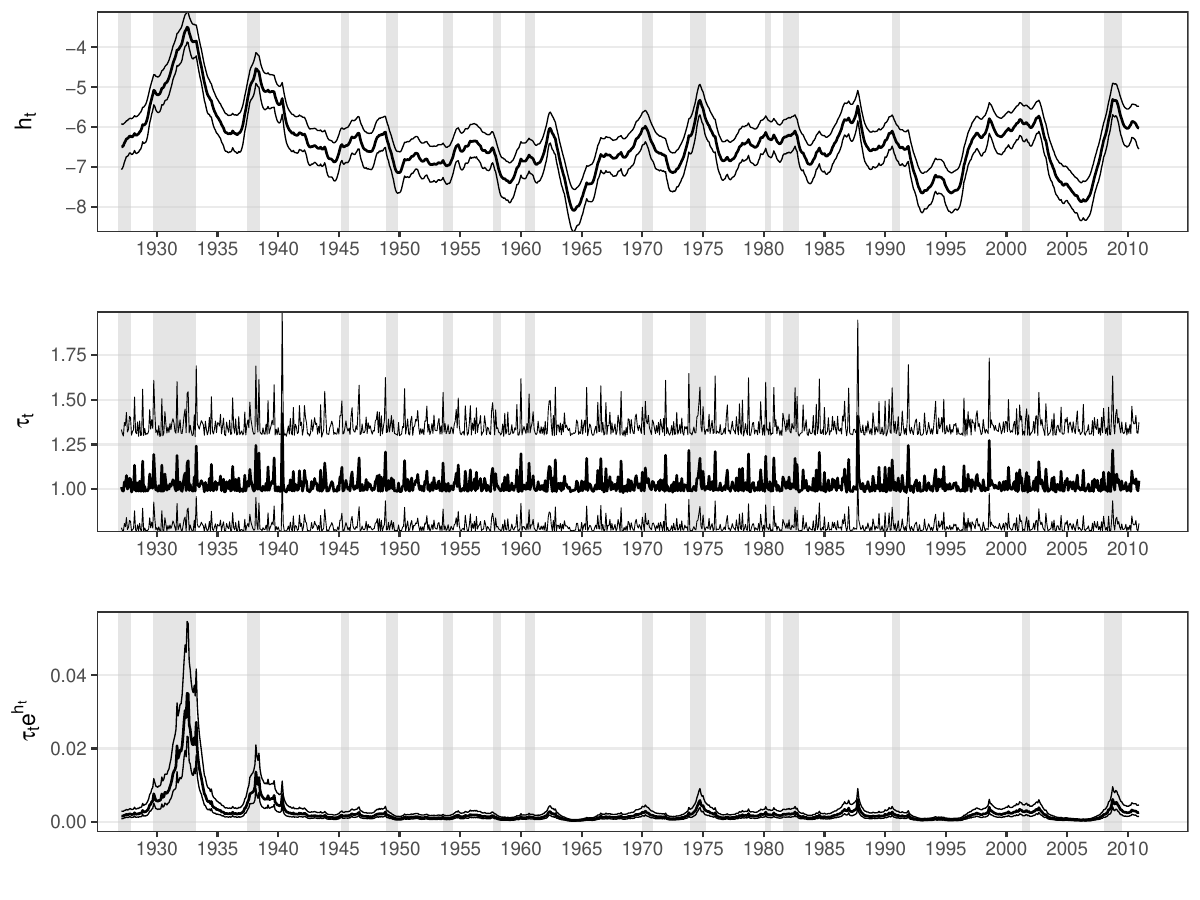}
\caption{Volatility related components over time: 1927:01 to 2010:12. Grey shaded areas indicate NBER recessions.}
\label{fig:volatilities}
\end{figure}

Figure~\ref{fig:volatilities} depicts the evolution of the three volatility components in the measurement equation: The upper panel shows the log-volatilities $h_t$ over time, the middle panel depicts the auxiliary scalings $\tau_t$ used to render the t-distribution conditionally Gaussian, and the bottom panel provides the combined series $\tau_t e^{h_t}$ of the measurement errors. The solid black line is the posterior median, the thin black lines indicate the 68 percent posterior coverage interval, while the grey shaded areas refer to National Bureau of Economic Research (NBER) recession dates. The sample starting in January 1927 features 15 distinct periods where the US economy was in recession, with an apparent empirical regularity that recessionary episodes are associated with elevated stock market volatility.

Volatilities in terms of the combined series in the bottom panel peak early in the sample during the Great Depression ranging from the end of 1929 to early 1933. The second largest peak occurs during the Recession of 1937--1938 which is usually considered minor to the Great Depression, even though it is among the worst recessions over the time span considered. For comparison, volatilities during this period reached levels almost twice as high than during the great financial crisis and the Great Recession from late 2007 to mid 2009. A further notable recessionary episode is the 1973 oil crisis coupled with the 1973--1974 stock market crash prominently discussed in the context of forecasting excess returns in \citet{welch2008comprehensive}. 

Apart from high-volatility episodes during recessions, some further stock market related events are worth mentioning. Figure \ref{fig:volatilities} clearly show the so-called Kennedy Slide of 1962, one of the first significant high-volatility periods after World War II, with large stock market declines. Moreover, the volatility series feature the famous Black Monday in October 1987, associated with the greatest one-day percentage decline in US stock market history. The Russian crisis and related collapse of the hedge fund Long-Term Capital Management in the late 1990s is visible, followed by a period of elevated volatilities prior to  the burst of the Dot-com bubble. An interesting observation is that such idiosyncratic events typically result in high frequency peaks in terms of $\tau_t$, indicating the necessity of a heavy-tailed error distribution to adequately adress such shocks.

\begin{figure}[t]
\includegraphics[width=\textwidth]{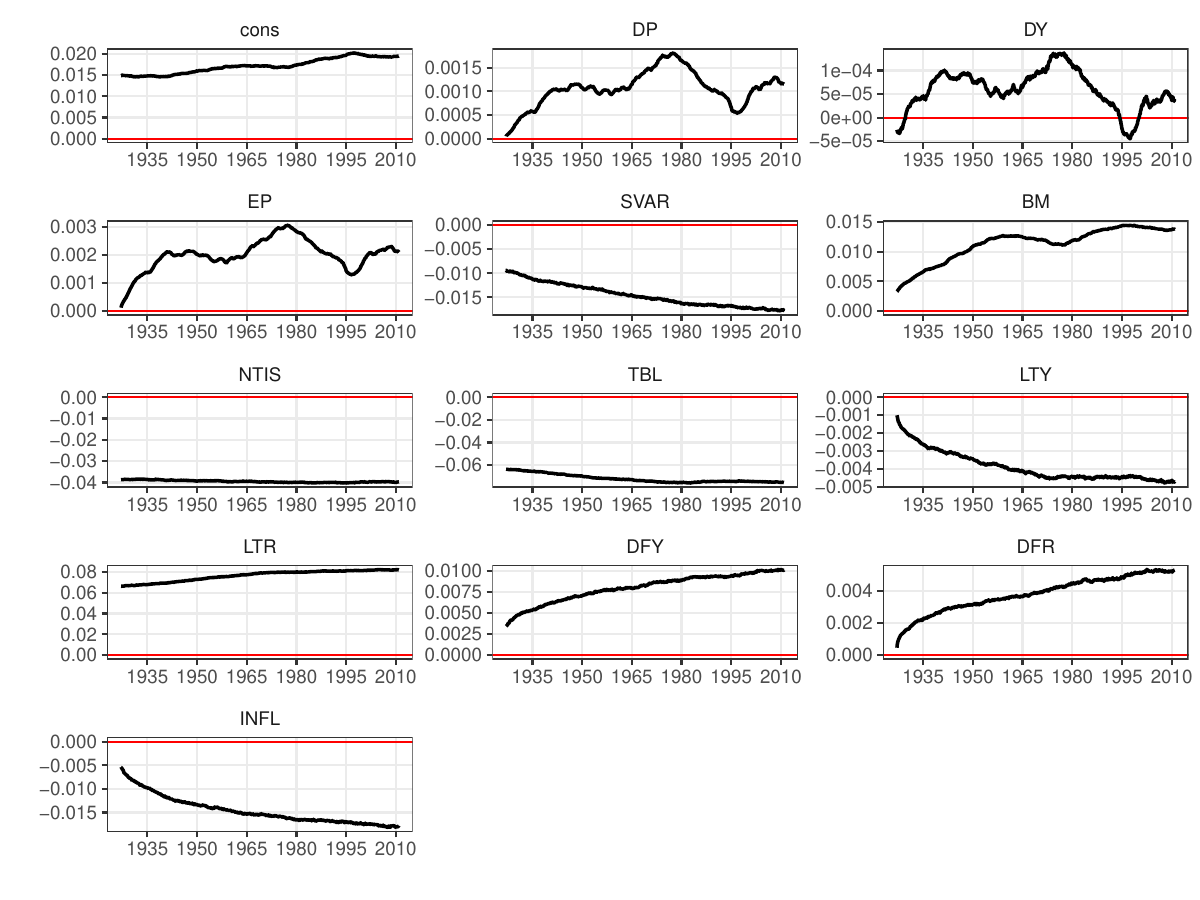}
\caption{Time-varying regression coefficients: 1927:01 to 2010:12.}
\label{fig:coefficients}
\end{figure}

We now turn to the time-varying regression coefficients associated with stock fundamentals, depicted in \autoref{fig:coefficients}. The solid black line indicates the posterior median,\footnote{We defer from showing 68 percent posterior coverage intervals since most of them cover zero, at least for some periods.} while the red line marks zero. We omit indicators for NBER recessions for better readability based on the notion that shifts in coefficients do not appear to be systematically related to distinct stages of the business cycle. The dynamic evolution of the series can be classified into three categories: First, some coefficients are approximately shrunk towards constancy. This class contains NTIS, TBL, LTR, and given the scale of the respective coefficient, also DY. Second, we obtain parameters that strictly decrease or increase over the sample period. Variables roughly featuring such coefficients are the intercept, SVAR, BM, LTY, DFY, DFR and INFL. Third, for DP and EP we observe coefficients of varying magnitude at different points in time. The paths of both states appear similar, with inital coefficients close to zero gradually gaining importance with peaks around 1980 and subsequent declines. Abrupt shifts governed by the t-distributed state equation errors mainly occur in the context of DP and EP.

The dynamics regarding the importance of covariates observed in \autoref{fig:coefficients} are roughly in line with the findings in \citet{welch2008comprehensive}, who estimate a set of models featuring different subsets of the variables and evaulate in-sample fit and out-of-sample forecast performance over time. Our study differs in the sense that the model includes all variables at once (labeled kitchen-sink regression in their study), and stochastically selects both inclusion and exclusion of quantities, besides whether variables differ in importance over time.

This concludes the section on in-sample empirical evidence. We proceed by discussing the design of the forecasting exercise and introduce a set of competing models for forecasting excess returns.

\subsection{Design of the forecasting exercise and competitors} \label{sec:design}
We utilize a recursive forecasting design and specify the period ranging from 1927:01 to 1956:12 as an initial estimation period. We then perpetually expand the initial estimation sample by one month until the end of the sample (2010:12) is reached. This yields a sequence of 647 monthly one-step-ahead predictive densities for S\&P 500 excess returns where we focus attention on root mean square forecast errors (RMSEs) and log predictive scores \citep[LPSs, see][for a discussion]{geweke2010comparing} to evaluate the predictive capabilities of the model. Compared to the existing literature \citep{lettau2001consumption, ang2007stock, welch2008comprehensive, dangl2012predictive}, this implies that we do not focus  on point predictions exclusively but rely on a more general measure that takes into account higher moments of the corresponding predictive densities.

Our set of competing models includes the historical mean with stochastic volatility (labeled Mean-SV). This model, a strong benchmark in the literature, enables us to evaluate whether the inclusion of additional explanatory variables improves forecasting. Moreover, we also include a constant parameter regression model with SV (referred to as Reg-SV), a recursive regression model (labeled Recursive), an autoregressive model of order one with SV (AR(1)-SV), a random walk without drift and with SV (RW-SV), and the mixture innovation model proposed in \cite{huber2017ttvp} featuring tresholded time-varying parameters (denoted TTVP). Moreover, to investigate which of the multiple features of our proposed model improve predictive capabilities, we include several nested versions: A time-varying parameter regression model with stochastic volatility and Gaussian shocks to both the measurement and the state equations with a DL shrinkage prior (labeled TVP-SV DL), a model that features t-distributed measurement errors (but Gaussian state innovations) and a DL prior (labeled t-TVP-SV DL 1), a specification that features t-distributed state innovations (but Gaussian measurement errors) and a DL prior (t-TVP-SV DL 2), and finally, the version of our proposed framework that features t-distributed state innovations and t-distributed measurement errors on top of the DL prior (t-TVP-SV DL 3).

A recent strand of the literature suggests that forecasts may be improved by selecting best-performing specifications dynamically from a pool of models, based on their past predictive performance \citep{raftery2010online,koop2012forecasting,onorante2016dynamic}. Such methods involve computing a set of weights $\mathfrak{w}_{t|t-1,m}$ at time $t$, conditional on information up to $t-1$ for each model $m$ within the model space $\mathcal{M}$. Specifically, we construct the weights as
\begin{equation}
\mathfrak{w}_{t|t-1,m} = \frac{\mathfrak{w}_{t-1|t-1,m}^\gamma}{\sum_{m\in\mathcal{M}}\mathfrak{w}_{t-1|t-1,m}^\gamma}, \quad \mathfrak{w}_{t-1|t-1,m} = \frac{\mathfrak{w}_{t-1|t-2,m}\times p_{t-1|t-2,m}}{\sum_{m\in\mathcal{M}}\mathfrak{w}_{t-1|t-2,m}\times p_{t-1|t-2,m}}.\label{eq:weights}
\end{equation}
Here, $p_{t-1|t-2,m}$ is the one-step ahead predictive likelihood, and the parameter $\gamma=0.99$ imposes persistence in the model weights over time. This parameter is a forgetting factor with values close to one yields a specification that takes into account also the less recent forecast performance. The initial model weights are assumed to be equal across all models. To choose the model per period, we select the one with the highest weight $\mathfrak{w}_{t|t-1,m}$ and label this approach dynamic model selection (DMS) in subsequent discussions.

\subsection{Predicting the US equity premium}\label{sec:forecast}
\autoref{tab:lps_results} displays relative RMSEs and differences in log predictive scores relative to the Mean-SV benchmark. For relative RMSEs, numbers exceeding unity indicate outperformance of the benchmark model, whereas numbers smaller than one indicate a stronger performance of the model under consideration. For the relative LPSs, a positive number indicates that a given model outperforms the benchmark model. We focus attention on forecasting accuracy during distinct stages of the business cycle (i.e.\ recessions/expansions), dated by the NBER Business Cycle Dating Committee. In doing so, we can investigate whether model performance changes over business cycle stages. Finally, we also report results over the full sample period.

\begin{table}[t]
\centering
\begin{tabular}{lccccccc}
  \toprule
  &\multicolumn{3}{c}{Relative root mean square errors} & &\multicolumn{3}{c}{Log Bayes factors} \\
  \cmidrule(l{3pt}r{3pt}){2-4} \cmidrule(l{3pt}r{3pt}){5-8}
 & Recession & Expansion & Full sample &  & Recession & Expansion & Full sample \\ 
  \midrule
Recursive & 0.914 & 0.945 & 0.933 &  & --- & --- & --- \\ 
  Reg-SV & 0.933 & 0.982 & 0.964 &  & 6.703 & 5.165 & 11.869 \\ 
  RW-SV & 1.006 & 1.000 & 1.000 &  & -0.426 & -3.124 & -3.550 \\ 
  AR(1)-SV & 0.970 & 0.969 & 0.968 &  & 3.124 & 7.315 & 10.439 \\ \midrule
  TVP-SV DL & 0.918 & 0.955 & 0.941 &  & 8.900 & 8.205 & 17.105 \\ 
  t-TVP-SV DL (1) & 0.923 & 0.955 & 0.943 &  & 8.416 & 9.627 & 18.043 \\ 
  t-TVP-SV DL (2) & 0.922 & 0.953 & 0.941 &  & 8.716 & 8.938 & 17.654 \\ 
  t-TVP-SV DL (3) & 0.929 & 0.960 & 0.948 &  & 8.103 & 3.231 & 11.334 \\ 
  TTVP & 0.945 & 0.966 & 0.957 &  & 6.074 & 8.293 & 14.367 \\ \midrule
  DMS & 0.925 & 0.955 & 0.944 &  & 8.170 & 12.066 & 20.236 \\ 
   \bottomrule
\end{tabular}
\caption{Root mean square errors and log predictive scores relative to the historical mean with SV model.}\label{tab:lps_results}
\end{table}

We start by considering point forecasting performance before turning to density forecasts. The left panel of \autoref{tab:lps_results} suggests that most specifications considered improve upon the Mean-SV benchmark over the full sample, as well as during recessionary and expansionary episodes. We find that the t-TVP-SV specifications with a DL prior all perform rather well, outperforming the benchmark up to over eight percent during recessions (in the case of the TVP-SV DL) and up to 5.7 percent over the full sample. It is noteworthy that constant parameter models, while outperforming the no-predictability benchmark, only yield small gains in predictive accuracy and this result confirms findings in \cite{welch2008comprehensive} and \cite{dangl2012predictive}. Interestingly, the recursive regression model outperforms all other specifications in terms of point forecasts, both in recessions and expansions, and over the full sample period. The DMS point forecasts are rather close to those of the time-varying parameter specifications.

One key finding is that accuracy improvements in recessions tend to be more pronounced, indicating that using more information seems to pay off during economic downturns.  We conjecture that larger information sets contain additional information necessary to better predict directional movements and this, in turn, improves point forecasting performance. Considering the results during expansions yields a similar picture: all state space models using some sort of shrinkage (including the TTVP specification) display a favorable point forecasting performance. While differences across models appear to be rather muted, this small premium in forecasting accuracy can be traced back to a feature attributed to the combination of shrinkage priors and heavy tailed process innovations.

The discussion above focused on point forecasts exclusively. To additionally assess how well the models perform in terms of density forecasting, the right panel of \autoref{tab:lps_results} presents relative LPSs. Note that the recursive regression model does not produce a full predictive distribution and is thus not included. A few results are worth emphasizing. First, dynamically selecting the best performing model over time based on past predictive likelihoods pays off and yields superior performance in terms of density forecasts for the full sample and expansions. Second, focusing on individual specifications over the model space, the last column of \autoref{tab:lps_results} reveals that most models under consideration outperform the historical mean model with SV by large margins over the full sample. This finding can be traced back to the fact that the Mean-SV includes no additional covariates and is thus unable to explain important features of the data that are effectively picked up by having additional exogenous covariates. Considering the forecast differences across models shows that introducing shrinkage in the TVP regression framework seems to pay off. Notice, however, that in terms of predictive capabilities, it suffices to allow for fat tailed innovations in either the state or measurement errors. Allowing for t-distributed errors for the shocks in the state and the observation equation generally yields weaker forecasting performance. A closer look at the underlying predictive density reveals that the predictive variance in that case appears to be slightly overestimated relative to the simpler specifications.

Second, zooming into the results for distinct stages of the business cycles indicates that t-TVP-SV DL 2 outperforms all competing model specifications during recessions. Especially when benchmarked against a simple random walk and the historical mean model, we find sharp increases in predictive accuracy when the more sophisticated approach is adopted. Considering the results for a constant parameter regression model also points towards favorable  predictive characteristics of this simple specification in terms of density predictions. As in the case of point forecasts, we generally attest our models more predictive capabilities during business cycle downturns and are thus in line with the recent literature \citep{rapach2010out, henkel2011time, dangl2012predictive}.

This result, however, does not carry over to expansionary stages of the business cycle. The penultimate column of \autoref{tab:lps_results} clearly shows that while models that perform well during recessions also tend to do well in expansions, the single best performing model is the t-TVP-SV DL 1 specification. By contrast, the flexible t-TVP-SV DL 3 model performs poorly during expansions. This stems from the fact that equity price growth appears to be quite stable during expansions and thus corroborates the statement above: in expansions, this specification simply yields inflated credible intervals and thus weaker predictive density forecasting performance.

These findings suggest that the strong overall performance of t-TVP-SV DL 1 is mainly driven by superior forecasting capabilities during expansions, whereas this model is slightly outperformed by t-TVP-SV DL 2 during recessionary periods. During turbulent times, we find that controlling for heteroscedasticity is important, corroborating findings reported in the literature \citep{clark2011real, clark2015macroeconomic, huber2016density, kas:spa}. Moreover, the results also indicate that allowing for heavy tailed shocks to the states helps capturing sudden shift in the regression coefficients, a feature that appears to be especially important during recessions.

\begin{figure}[t]
\includegraphics[width=\textwidth]{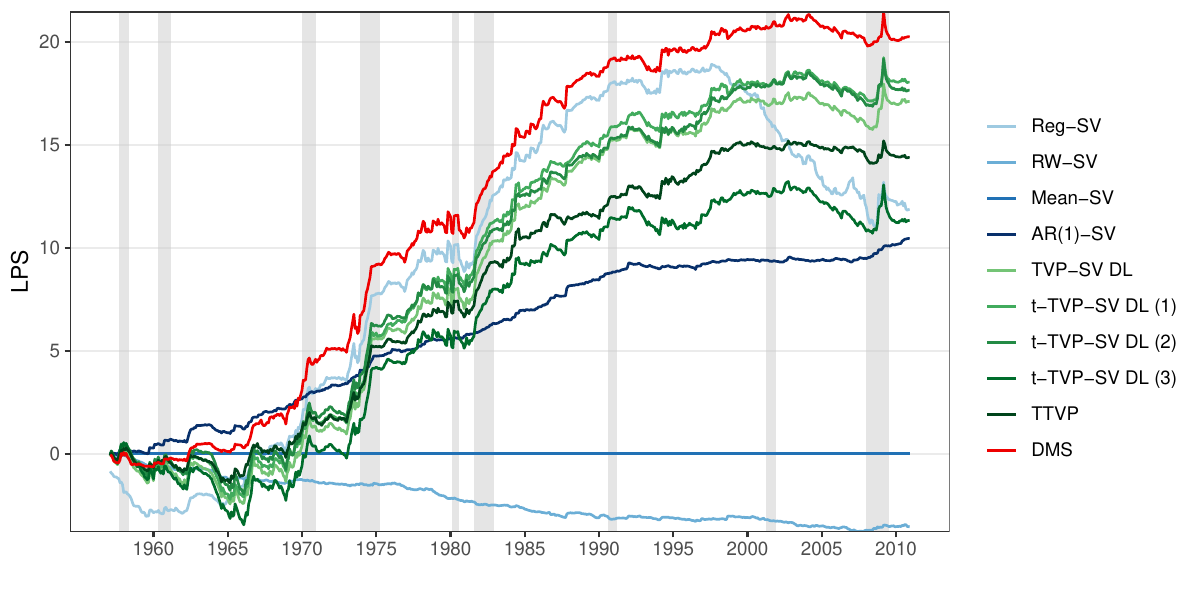}
\caption{Log predictive Bayes factors relative to the historical mean model with stochastic volatility: 1957:01 to 2010:12. Grey shaded areas indicate NBER recessions.}
\label{fig:lps_overtime}
\end{figure}

\begin{figure}[t]
\includegraphics[width=\textwidth]{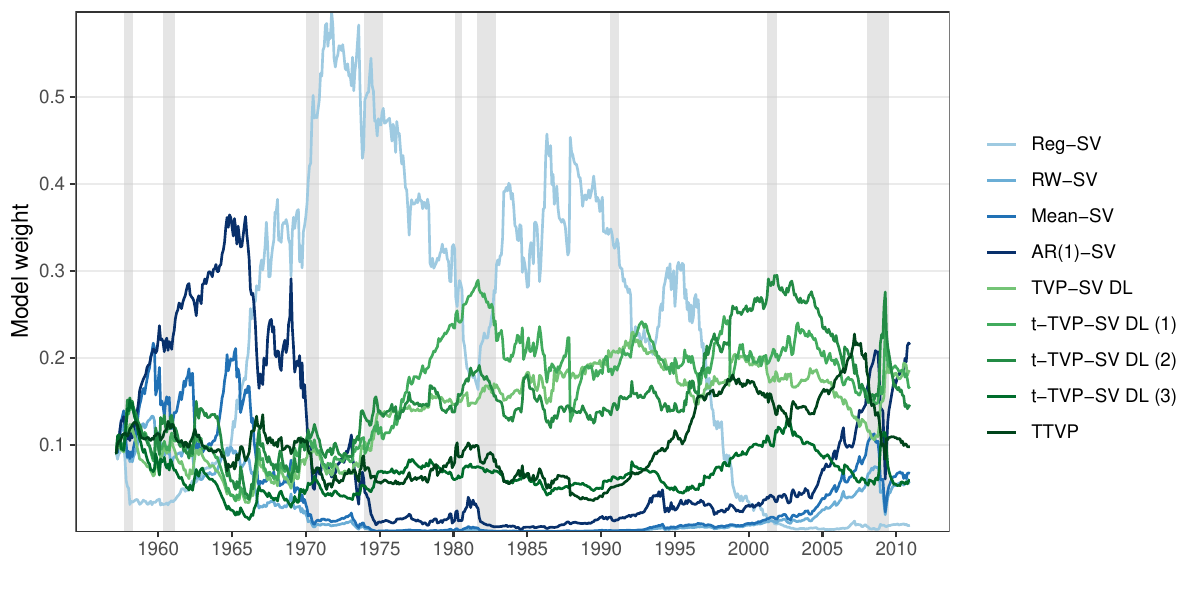}
\caption{Dynamic model weights $\mathfrak{w}_{t|t-1,m}$ for model selection: 1957:01 to 2010:12. Grey shaded areas indicate NBER recessions.}
\label{fig:weights_overtime}
\end{figure}

The previous discussion focused on overall forecast performance and highlighted that predictive accuracy depends on the prevailing economic regime. In crisis episodes, models that are generally quite flexible yield pronounced accuracy increases. Moreover, there is substantial evidence for predictive gains when dynamically selecting models. In the next step, we analyze whether there exists additional heterogeneity of forecast performance over time that is not specific whether the economy is in a recession or expansion. To this end, \autoref{fig:lps_overtime} displays the evolution of the relative LPSs over time, and \autoref{fig:weights_overtime} relatedly indicates the underlying model weights for the DMS specification. 

Figure~\ref{fig:lps_overtime} indicates that the DMS specification outperforms all other specifications for the most part of the holdout sample. This implies that the approach to calculating model weights appears to capture shifts in a model's predictive performance quite well. After an initial period from the start of the holdout to the beginning of the 1970s, the AR(1)-SV specification is the best performing model. From the midst of the 1970s up to the midst of the 1990s, a constant parameter model with SV outperformed all models considered. From around 1995 onwards, we observe a pronounced decline in forecasting performance of the Reg-SV specification over time while all models that feature time-variation in their parameters produced a rather stable predictive performance. During the great financial crisis, all models except the RW-SV outperform the benchmark. This again highlights that especially during crisis episodes, introducing shrinkage and time-varying parameters yields pronounced gains in forecast accuracy.

Given the specification of $\mathfrak{w}_{t|t-1,m}$ in \autoref{eq:weights} and the evolution of LPSs in \autoref{fig:lps_overtime}, the findings for the model weights depicted over time in \autoref{fig:weights_overtime} are unsurprising. After an initial eight-year period where the proposed procedure dynamically selected the AR(1)-SV model, the dominating model until 1980 is the regression model with constant parameters and SV. Subsequently, for a brief period of approximately three years, t-TVP-SV-DL 1 received the largest model weight. Afterwards, up to the mid/late 1990s, the constant parameter model with SV, again, was selected as the best-performing model based on past predictive likelihoods. The pronounced decline in forecast performance discussed for Reg-SV in the context \autoref{fig:lps_overtime}, however, also resulted in the model essentially receiving zero weight from 1995 onwards, where t-TVP-SV-DL 2 receives the highest weights in most cases.

\begin{figure}[t]
    \centering
    \begin{subfigure}[t]{0.5\textwidth}
        \centering
        \includegraphics[width=\textwidth]{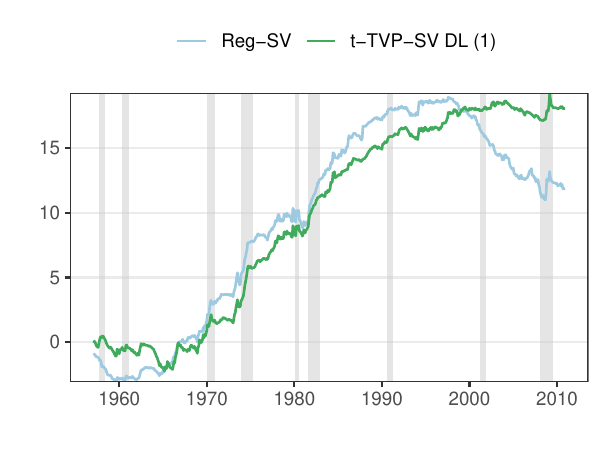}
        \caption{Log predictive Bayes factors}
    \end{subfigure}%
    ~ 
    \begin{subfigure}[t]{0.5\textwidth}
        \centering
        \includegraphics[width=\textwidth]{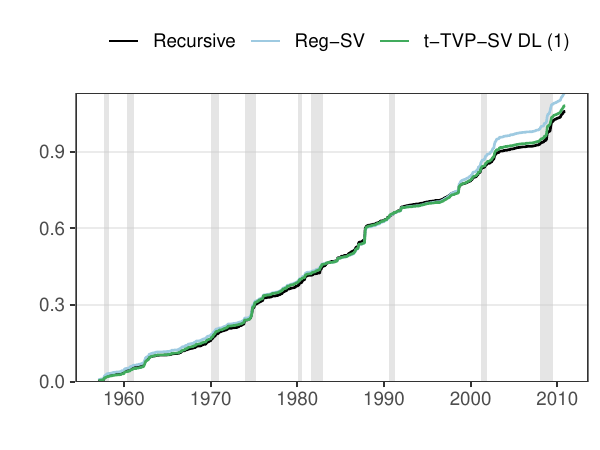}
        \caption{Cumulative squared forecast errors}
    \end{subfigure}
\caption{Log predictive Bayes factors relative to the historical mean model and cumulative squared forecast errors: Best performing models. Grey shaded areas indicate NBER recessions.}
\label{fig:topmodels}
\end{figure}

In order to investigate where forecasting gains stem from, \autoref{fig:topmodels}, left panel, displays the log predictive Bayes factors of Reg-SV and t-TVP-SV DL 1 relative to Mean-SV, whereas the right panel shows the cumulative squared forecast errors over the hold-out period. This figure clearly suggests that the sharp decline in predictive accuracy of the Reg-SV model mainly stems from larger forecast errors as opposed to other features of the predictive density. The weaker point forecasting performance can be explained by the lack of time-variation in the parameters of the Reg-SV model. Notice that the recursive forecasting design implies that coefficients are allowed to vary over the hold-out period but comparatively slower as under a time-varying parameter regression framework. Thus, while the coefficients in t-TVP-SV DL 1 are allowed to change rapidly if economic conditions change, the coefficients in Reg-SV take longer to adjust and this might be detrimental for predictive accuracy. For the sake of completeness, we also include the recursive regression in the right panel. An interesting finding is that homoscedastic errors appear to result in lower squared forecast errors, comparable to those of t-TVP-SV DL 1.

\section{Concluding remarks}
This paper proposes a flexible econometric model that introduces shrinkage in the general state space modeling framework. We depart from the literature by assuming that the shocks to the state as well as observation are potentially non-Gaussian and follow a t-distribution. Assuming heavy tailed measurement errors allows to capture outlying observations, while t-distributed errors in the state equation allow for large shocks to the latent states. This feature, in combination with a set of global-local shrinkage priors, allows for flexibly assessing whether time-variation is necessary and also, to a certain extent, mimics the behavior of models with a low number of potential regime shifts.

In the empirical application we forecast S\&P~500 excess returns. Using a  panel of macroeconomic and financial fundamentals and a large set of competing models that are commonly used in the literature, we show that our proposed modeling framework yields sizeable gains in predictive accuracy, both in terms of point and density forecasting.  We find that using the most flexible specification generally does not pay off relative to using a somewhat simpler specification that either assumes t-distributed shocks in the measurement errors or in the state innovations. Especially during economic downturns, we find that combining shrinkage with non-Gaussian features in the state equation yields strong point and density predictions whereas in expansions, a model with t-distributed measurement errors performs best. This model also performs best if the full hold-out period is taken into consideration.



\normalsize
\singlespacing
\bibliographystyle{./bibtex/fischer}
\bibliography{./bibtex/ref}
\addcontentsline{toc}{section}{References}
\newpage

\begin{appendices}

\end{appendices}

\end{document}